\documentclass[reprint, twocolumn, superscriptaddress]{revtex4-2}

\usepackage[utf8]{inputenc}
\usepackage{graphicx}
\usepackage{wrapfig}
\usepackage[colorlinks,allcolors=black,citecolor=blue,urlcolor=blue]{hyperref}
\usepackage{chemformula}
\usepackage{exsheets}
\usepackage{paralist}
\usepackage{tcolorbox}
\usepackage{bm}
\usepackage{physics}
\usepackage{xcolor}
\usepackage{cancel}
\usepackage{graphicx}
\usepackage{dcolumn}
\usepackage{filecontents}
\usepackage{soul}
\usepackage{breakurl}

\renewcommand{\figurename}{\textbf{Fig.}}
\renewcommand{\tablename}{\textbf{Table}}
\makeatletter
\renewcommand{\fnum@figure}{\textbf{\figurename~\thefigure}}
\renewcommand{\fnum@table}{\textbf{\tablename~\thetable}}
\makeatother

\begin{document}
	
	\title{Quasi-one-dimensional hydrogen bonding in nanoconfined ice}
	
	\author{Pavan Ravindra}
	\thanks{
		These authors contributed equally.	
	}
	\affiliation{%
		Yusuf Hamied Department of Chemistry, University of Cambridge, Lensfield Road, Cambridge, CB2 1EW, UK
	}
	\affiliation{%
		Department of Chemistry, Columbia University, 3000 Broadway, New York, NY 10027, USA
	}
	
	\author{Xavier R. Advincula}
	\thanks{
		These authors contributed equally.	
	}
	\affiliation{%
		Yusuf Hamied Department of Chemistry, University of Cambridge, Lensfield Road, Cambridge, CB2 1EW, UK
	}
	\affiliation{%
		Cavendish Laboratory, Department of Physics, University of Cambridge, Cambridge, CB3 0HE, UK
	}
	\affiliation{%
		Lennard-Jones Centre, University of Cambridge, Trinity Ln, Cambridge, CB2 1TN, UK
	}
	
	\author{Christoph Schran}%
	\affiliation{%
		Cavendish Laboratory, Department of Physics, University of Cambridge, Cambridge, CB3 0HE, UK
	}
	\affiliation{%
		Lennard-Jones Centre, University of Cambridge, Trinity Ln, Cambridge, CB2 1TN, UK
	}
	
	\author{\\Angelos Michaelides}%
	\email{am452@cam.ac.uk}
	\affiliation{%
		Yusuf Hamied Department of Chemistry, University of Cambridge, Lensfield Road, Cambridge, CB2 1EW, UK
	}
	\affiliation{%
		Lennard-Jones Centre, University of Cambridge, Trinity Ln, Cambridge, CB2 1TN, UK
	}
	
	\author{Venkat Kapil}%
	\email{vk380@cam.ac.uk}
	\affiliation{%
		Yusuf Hamied Department of Chemistry, University of Cambridge, Lensfield Road, Cambridge, CB2 1EW, UK
	}
	\affiliation{%
		Lennard-Jones Centre, University of Cambridge, Trinity Ln, Cambridge, CB2 1TN, UK
	}
	\affiliation{%
		Department of Physics and Astronomy, University College London, 17-19 Gordon St, London WC1H 0AH, UK
	}
	\affiliation{%
		Thomas Young Centre and London Centre for Nanotechnology, 19 Gordon St, London WC1H 0AH, UK
	}
	
	\begin{abstract}
		
		\noindent
		The Bernal–Fowler ice rules stipulate that each water molecule in an ice crystal should form four hydrogen bonds.
		However, in extreme or constrained conditions, the arrangement of water molecules deviates from conventional ice rules, resulting in properties significantly different from bulk water. 
		In this study, we employ machine learning-driven first-principles simulations to identify a new stabilization mechanism in nanoconfined ice phases beyond conventional ice rules.
		Instead of forming four hydrogen bonds, nanoconfined crystalline ice can form a quasi-one-dimensional hydrogen-bonded structure that exhibits only two hydrogen bonds per water molecule. 
		These structures consist of strongly hydrogen-bonded linear chains of water molecules that zig-zag along one dimension, stabilized by van der Waals interactions that stack these chains along the other dimension.
		The unusual interplay of hydrogen bonding and van der Waals interactions in nanoconfined ice results in atypical proton behavior such as potential ferroelectric behavior, low dielectric response, and long-range proton dynamics.

	\end{abstract}
	
	\maketitle
	
	\clearpage
	\newpage
	\mbox{~}
	\clearpage
	\newpage
	
	\section{Introduction}

	Nanoconfined water is prevalent in biological and geological systems, and it has technological applications in tribology~\cite{kenis_microfabrication_1999}, desalination~\cite{surwade_water_2015}, and clean energy~\cite{kalra_osmotic_2003}.
	Experiments have revealed a diverse range of unusual properties in these systems containing confined water~\cite{lizee2023strong,canale2019nanorheology}.
	While a unified theory of the thermodynamics of systems confined to nanoscale dimensions is missing~\cite{coquinot2023quantum,faucher_critical_2019}, atomistic simulations have aided our interpretation of experiments by revealing the microscopic structure and dynamics of nanoscale water~\cite{truskett_thermodynamic_2001, zhao_highly_2014, chakraborty_confined_2017, li_replica_2019}.
	Specifically, first-principles simulations provide a bottom-up format for predicting the physicochemical behavior of water in complex conditions~\cite{corsetti_structural_2016, nagata2015ultrafast, chen_two_2016, jiang_first-principles_2021, kapil_first-principles_2022, lin_temperature-pressure_2023} where empirical models fitted to bulk data may be insufficient.  \\

	Recent first-principles studies, direct~\cite{jiang_first-principles_2021} or machine-learning-accelerated~\cite{kapil_first-principles_2022, lin_temperature-pressure_2023}, have explored finite-temperature phase behaviors of nanoconfined water in atomistically-flat nanocavities -- meant to resemble the conditions of water trapped between two graphene sheets.
	Notably, Kapil et al.~\cite{kapil_first-principles_2022} calculated the temperature-pressure phase diagram of a monolayer film of water by using machine learning potentials (MLPs)~\cite{behler_constructing_2015, schran_committee_2020}.
	They unveiled crystalline monolayer phases comprising hexagonal, pentagonal, and rhombic motifs, an entropically stabilized hexatic phase of water, and a superionic phase at dramatically lower temperatures and pressures than its bulk counterpart~\cite{sun_phase_2015, millot_experimental_2018, millot_nanosecond_2019, cheng_phase_2021}. 
	More recently, Lin et al.~\cite{lin_temperature-pressure_2023} extended this investigation to larger confinement widths and a wider range of confinement pressures, revealing new bilayer ice phases. 
	An intriguing aspect of these nanoconfined ice phases is their deviation from the Gibbs-Thomson relation~\cite{kapil_first-principles_2022, lin_temperature-pressure_2023}.
	These phases aren't simply bulk ice phases with altered phase boundaries; rather, they consist of unique motifs and topologies that are not observed in bulk ice. \\

	Given the significant differences in the structure of bulk and nanoconfined water, it is essential to understand the underlying principles that distinguish their phase behaviors.
	The behaviors of crystalline ice are well explained using Pauling's principle of hydrogen bonding ~\cite{pauling_nature_1931} and the Bernal-Fowler ice rules, which state that each water molecule should participate in four total hydrogen bonds~\cite{bernal1933theory}.
	However, these ice rules can be broken under extreme conditions and through translational symmetry disruptions at interfaces and in nanoscale confinement.
	These violations result in a variety of unusual phase behaviors, including a quantum delocalized state of water across nanocapillaries~\cite{kolesnikov_quantum_2016}, a square-like ice phase encapsulated between graphene sheets~\cite{algara-siller_algara-siller_2015}, thin epitaxial films of ice exhibiting ferroelectric nature~\cite{sugimoto_emergent_2016}, and the formation of water chains~\cite{forster_water-hydroxyl_2011} and nanoclusters resembling cyclic hydrocarbons on metals~\cite{liriano_waterice_2017}.
	Attempts have been made to extend the Bernal-Fowler ice rules to these settings, and such modified ice rules generally focus on incorporating water-surface interactions into the hydrogen bonding picture of the Bernal-Fowler ice rules~\cite{cerda_novel_2004, corsetti_structural_2016, liriano_waterice_2017}.
	However, several recently predicted nanoconfined ice phases exhibit fewer than four hydrogen bonds per water molecule~\cite{jiang_first-principles_2021, kapil_first-principles_2022, lin_temperature-pressure_2023}, which brings into question the extent to which nanoconfined water phases are stabilized by hydrogen bonding and if such modified ice rules should be centered around hydrogen bonding at all. \\

	In this work, we utilize first-principles calculations and MLPs to identify a new stabilization mechanism in nanoconfined ice phases beyond conventional ice rules.
	Instead of forming four hydrogen bonds per water molecule, nanoconfined ice phases can form zig-zagging quasi-one-dimensional hydrogen-bonded chains stacked together by van der Waals (vdW) interactions.
	We identify this mechanism across monolayer and bilayer crystalline ice found in Refs.~\citenum{jiang_first-principles_2021, kapil_first-principles_2022, lin_temperature-pressure_2023} in confinement slits of varying sizes.
	Employing MLPs to access long timescales at significantly lower computational costs than traditional first-principles methods, we show that quasi-one-dimensional hydrogen bonding is stable at finite temperatures up to the melting point of nanoconfined ice.
	The arrangement of strong and weak interactions in nanoconfined water bears similarities to two-dimensional vdW materials consisting of so-called quasi-one-dimensional chains~\cite{li2020diverse,balandin2022one}.
	Conventionally, these materials comprise strong covalent bonds within the quasi-one-dimensional chains (as opposed to hydrogen bonds), with weak inter-chain vdW attraction holding these chains together.
	These materials exhibit unusual electronic, vibrational, and optical properties~\cite{li2020diverse,balandin2022one}, potentially exhibiting superconductivity at higher pressures~\cite{zhang2019new}.
	Analogously, the intrinsic anisotropic bonding in the flat-rhombic phase leads to anomalous rotational dynamics of water molecules, including detectable long-range concerted disorder on the scale of nanometers, which has potential applications for molecular devices.

	\begin{table*}[!t]
		\caption{\textbf{$\vert$ First-principles calculations for four nanoconfined ice phases.} Average number of hydrogen bonds and energies of the hexagonal phase at 0.1\,GPa, the pentagonal phase at 0.3\,GPa, the flat-rhombic phase at 2.0\,GPa, and the ZZ-qBI phase at 20\,GPa in their equilibrium structures. We report the average number of donated, accepted, and total hydrogen bonds per water molecule. The energy values include the total and vdW contributions to the lattice energy $E_{\text{lattice}}$, the cohesive energy of a quasi-one-dimensional chain of water molecules $E_{\text{chain}}$, and the remaining stabilization energy from interactions between chains of water molecules $E_{\text{stack}}$. The equilibrium structures of the ice phases were obtained from fixed-cell geometry optimizations, as explained in the methods section. The energies were obtained directly from first-principles calculations. Hydrogen bonds were detected using the standard geometric criteria from Ref.~\citenum{luzar_hydrogen-bond_1996}. For the hexagonal and pentagonal phases, we report the average value and the standard deviation as the error bar, where the standard deviation is computed across different choices of chains, as described in the text.}
		\begin{ruledtabular}
			\begin{tabular}{l || c | c | c || c | c || c | c || c | c }
				Phase & \multicolumn{3}{c||}{Geometric hydrogen bonds} &  \multicolumn{2}{c||}{$E_{\text{lattice}}$ (kcal/mol)} & \multicolumn{2}{c||}{$E_{\text{chain}}$ (kcal/mol)} & \multicolumn{2}{c}{$E_{\text{stack}}$ (kcal/mol)} \\
				\hline
				& donors & acceptors  & total & total & vdW & total & vdW & total & vdW \\
				\hline
				Hexagonal & 1.5 & 1.5 & 3.0 & -9.05 & -1.79 & -5.12 $\pm$ 0.31 & -0.85 $\pm$ 0.17 & -3.93 $\pm$ 0.31 & -0.94 $\pm$ 0.17 \\
				Pentagonal & 1.6 & 1.6 & 3.2 & -8.99 & -2.24 & -3.20 $\pm$ 0.50 & -1.10 $\pm$ 0.50 & -5.80 $\pm$ 0.50 & -1.10 $\pm$ 0.50 \\
				Flat-rhombic & 1.0 & 1.0 & 2.0 & -8.75 & -2.76 & -6.76 & -1.08 & -2.00 & -1.68 \\
				ZZ-qBI & 1.0 & 1.0 & 2.0 & -12.71 & -19.91 & -9.69 & -4.65 & -3.05 & -15.33 \\
			\end{tabular}
		\end{ruledtabular}
		\label{tab:ice_phases}
	\end{table*}

	\section{Results}
	
	\subsection{Violations of conventional ice rules lead to a quasi-one-dimensional hydrogen bonded ice phase}

	In bulk ice, the Bernal-Fowler ice rules dictate that each water molecule in a stable ice phase should participate in four hydrogen bonds with its neighboring water molecules: 2 donated and 2 accepted hydrogen bonds~\cite{bernal1933theory}.
	Previous work on nanoconfined water at the DFT level has suggested that the thermodynamically stable monolayer phase is a square structure that still exhibits this 4-fold coordination ~\cite{zhao_highly_2014, corsetti_structural_2016}.
	However, more recent first-principles-level simulations have suggested a wide variety of stable phases of nanoconfined water spanning a broad range of different temperature and lateral pressure conditions~\cite{chen_two_2016,jiang_first-principles_2021, kapil_first-principles_2022,lin_temperature-pressure_2023}.
	Kapil et al.~\cite{kapil_first-principles_2022} report a hexagonal phase stable below $0.1$\,GPa, a pentagonal phase stable from $0.1-0.5$\,GPa, and a flat-rhombic phase stable above $0.5$\,GPa for a confinement width of 5\,\AA{}.
	Ref.~\citenum{kapil_first-principles_2022} also confirms the greater thermodynamic stabilities of these phases with respect to the square phase using the chemically accurate and robust Quantum Monte Carlo (QMC) calculation setup~\cite{zen_boosting_2016, della_pia_dmc-ice13_2022}.
	Refs.~\citenum{jiang_first-principles_2021} and~\citenum{lin_temperature-pressure_2023} explored even higher lateral pressures for a confinement width of 6\,\AA{}.
	These studies found that the stable phase beyond 0.5 GPa is a zigzag monolayer ice (ZZMI) phase, which is topologically equivalent to the flat-rhombic phase, except that the larger confinement width leads to a small buckling of the oxygen atoms.
	In the 15-20\,GPa regime, Lin et al. report a zigzag quasi bilayer ice (ZZ-qBI) structure as the stable phase~\cite{lin_temperature-pressure_2023}.
	This phase also resembles the flat-rhombic phase, although the buckling in this phase results in two nearly distinct layers of water molecules.
	\\
	
	\begin{figure*}
		\includegraphics{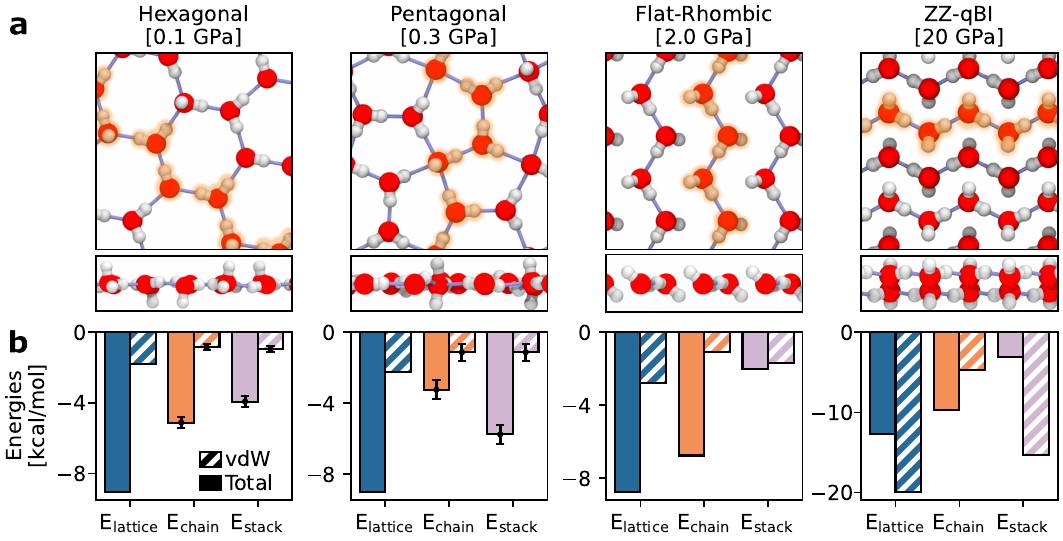}
		\caption{\textbf{$\vert$ Nanoconfined ice phases and the role of vdW stacking of hydrogen bonded chains. } a) The four nanoconfined ice phases considered in this work from Ref.~\citenum{kapil_first-principles_2022} and Ref.~\citenum{lin_temperature-pressure_2023}, along with the lateral pressures used for simulating each phase. Solid lines between water molecules indicate hydrogen bonds. We highlight chains of hydrogen-bonded water molecules in each phase. For estimating $E_{\mathrm{chain}}$, we use the unique highlighted chains in the case of the flat-rhombic and ZZ-qBI phases. For the hexagonal and pentagonal phases, we report the average value and the standard deviation as the error bar, where the standard deviation is computed across different choices of chains, as described in the text. b) For each phase, we compute the stabilization energy of a single molecule in the 0\,K crystal structure, $E_{\mathrm{lattice}}$; the (average) stabilization energy of a water molecule within the hydrogen bonded chain(s), $E_{\mathrm{chain}}$; and the stabilization energy between chains in the crystal structure, $E_{\mathrm{stack}}$. The hatched bars show the contribution of vdW interactions to the corresponding stabilization energy. Source data are provided as a Source Data file.}
		\label{fig:fig1}
	\end{figure*}

	As shown in Fig.~\ref{fig:fig1}(a) and Table~\ref{tab:ice_phases}, none of the above thermodynamically stable nanoconfined ice phases satisfy conventional ice rules: all of these phases display fewer than four hydrogen bonds per water molecule. 
	However, the water molecules in each of these phases violate the ice rules in different ways.
	In the low-pressure hexagonal phase, all water molecules exhibit a 3-fold hydrogen bonding coordination: half of the water molecules donate two and accept one hydrogen bond, and the other half donate one and accept two.
	Water molecules in the intermediate-pressure pentagonal phase exhibit diverse hydrogen bonding motifs, with coordination numbers ranging from 2 to 4. 
	This results in an average hydrogen bond coordination of 3.2 in the pentagonal crystal structure.
	The high-pressure flat-rhombic (or the ZZMI) phase exhibits a single hydrogen-bonding motif in which each water molecule donates one and accepts one hydrogen bond.
	The resulting hydrogen bonding network contains just two hydrogen bonds per water molecule, deviating significantly from the conventional ice rules.
	The ZZ-qBI phase exhibits a similar network, with just two hydrogen bonds per water molecule.
	Since this phase is stable even at lateral pressures much higher than the flat-rhombic phase, this suggests that such extreme violations of the bulk ice rules are likely to occur in high lateral pressure conditions.
	\\

	The flat-rhombic phase is thermodynamically stable in a sizeable temperature-pressure region of the phase diagram~\cite{jiang_first-principles_2021,kapil_first-principles_2022, lin_temperature-pressure_2023} despite having just two hydrogen bonds per molecule. 
	Considering that ice in general is primarily stabilized by a dense network of hydrogen bonds~\cite{pauling_nature_1931, pauling1935structure}, this stability over a broad range of conditions is highly unusual.
	To understand this unexpected behavior, we examine the structure of the flat-rhombic phase and look for similarities with other types of materials.
	As depicted in Figure~\ref{fig:fig1}(a), the flat-rhombic phase is characterized by a zigzagging hydrogen bond network that runs along quasi-one-dimensional chains~\cite{jiang_first-principles_2021,lin_temperature-pressure_2023}.
	These chains are stacked alongside each other without any hydrogen bonding between chains. 
	This structure is similar to quasi-one-dimensional vdW materials~\cite{li2020diverse,balandin2022one}, which exhibit unusual electronic, optical, and vibrational properties.
	However, while these materials consist of covalently bonded chains held together by weak inter-chain vdW interactions, the chains in the flat-rhombic phase are stabilized by hydrogen bonding. \\

	To determine whether the flat-rhombic phase can be classified as a quasi-one-dimensional vdW material, we calculate the binding energy of the one-dimensional chains and study the role of vdW interactions in stabilizing these chains. 
	In Fig.~\ref{fig:fig1}(b) and Table~\ref{tab:ice_phases}, we report three different types of binding energies of a water molecule in each phase: the stabilization energy of a water molecule in the corresponding crystal structure, $E_{\mathrm{lattice}}$; the stabilization energy of a water molecule within a continuous chain of hydrogen-bonded water molecules, $E_{\mathrm{chain}}$ (see highlighted chains of water molecules in Fig.~\ref{fig:fig1}(a)); and the remaining stabilization energy from interactions between chains of water molecules in each phase, $E_{\mathrm{stack}}$.
	This means that the stacking energy is the stabilization energy of the lattice that is not captured by the chains alone: $E_{\textrm{stack}} = E_{\textrm{lattice}}  - E_{\textrm{chain}}$.
	In the hexagonal phase, we consider two distinct chain types in the `zigzag' and `armchair' directions and report the average binding energies.
	As can be seen in the pentagonal phase snapshot in Fig.~\ref{fig:fig1}(a), there is no clear choice for a unique hydrogen bond chain in the pentagonal phase.
	This shows qualitatively that the pentagonal phase cannot clearly be separated into distinct hydrogen bond chains.
	Nonetheless, to perform the quantitative comparison shown in Fig.~\ref{fig:fig1}(b), we averaged our energies over three different hydrogen bond chains in the pentagonal phase.
	The energies for these different chains were all extremely close to each other, suggesting that this quantitative comparison is robust with respect to the exact choice of hydrogen bond chain in the pentagonal phase.
	For the flat-rhombic and ZZ-qBI phases, we considered a single unique chain. \\

	In Fig.~\ref{fig:fig1}(b), we observe that $E_{\mathrm{chain}}$ contributes to approximately 50\% or less of the crystal stabilization energy in both hexagonal and pentagonal phases. 
	Therefore, these phases have high stacking energy of chains, stabilized primarily by hydrogen bonds, with vdW interactions (indicated by hatched regions) playing only a secondary role in stabilizing the chains. 
	Conversely, in the flat-rhombic phase, the primary crystal stabilization source is from the chains. 
	Here, the stacking energy is low, with roughly 90\% of the stacking energy resulting from vdW interactions. 
	This difference indicates that the interactions stabilizing water molecules in the flat-rhombic phase contrast starkly with those in the hexagonal and pentagonal phases. 
	The flat-rhombic phase appears to form hydrogen bonds in one dimension while being stabilized by weak vdW forces in the other, classifying it as a quasi-one-dimensional vdW crystal. 
	Similarly, the stacking energy of the ZZ-qBI phase is almost entirely driven by vdW interactions, meaning that vdW forces also stabilize the interactions between this phase's hydrogen-bonded chains.
	In fact, in the absence of vdW stacking, the lattice energy is positive, meaning that the ZZ-qBI crystal structure would be thermodynamically unstable.
	This suggests that these vdW interactions between hydrogen-bonded chains play an essential role in stabilizing nanoconfined crystal structures at high lateral pressures.
	As our machine learning potential isn't trained for pressures beyond 8\,GPa or just revPBE0 level, we are unable to comment on the dynamical stability or metastability of the ZZ-qBI phase in the absence of the vdW interactions.
	
	\subsection{Anomalous finite-temperature dependence of hydrogen bonding}
	
	\begin{figure}
		\includegraphics{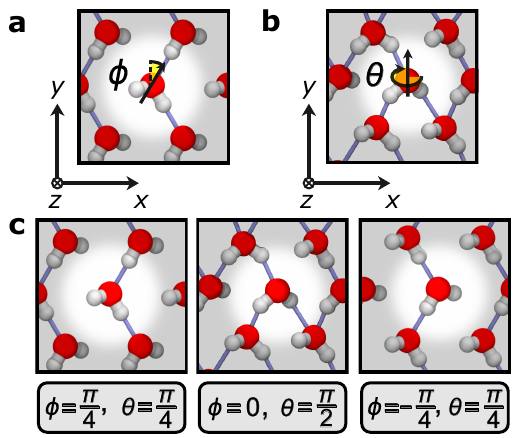}
		\caption{\textbf{$\vert$ Depiction of the orientations of water molecules.} a) The $\phi$ angle captures in-plane and out-of-plane fluctuations of water molecules' dipole moments. b) The $\theta$ angle captures in-plane rotations of water molecules. We refer the readers to section II.B for the mathematical definitions of the $\phi$ and $\theta$ angles. c) The left and right images depict the molecular orientations at the probability maxima in Fig.~\ref{fig:fig2}(a), while the middle image depicts the molecular orientation at the maximum for the number of putative hydrogen bonds in Fig.~\ref{fig:fig2}(b).}
		\label{fig:angle_viz}
	\end{figure}
	
	\begin{figure}
		\includegraphics{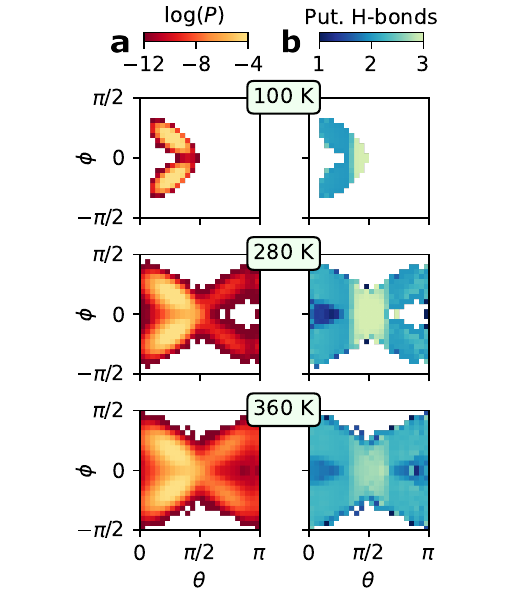}
		\caption{\textbf{$\vert$ Molecular orientation and the number of putative hydrogen bonds.} a) The two-dimensional log probabilities along the $\phi$ and $\theta$ parameters for the flat-rhombic phase at 2 GPa and three different temperatures. b) The average number of putative hydrogen bonds associated with each $(\theta,\phi)$ molecular orientation for the flat-rhombic phase under the same conditions. Relevant molecular orientations are depicted in Fig.~\ref{fig:angle_viz}(c). We employ the geometric definition of the hydrogen bond from Ref.~\citenum{luzar_hydrogen-bond_1996}. Source data are provided as a Source Data file.}
		\label{fig:fig2}
	\end{figure}
	
	The characteristic interactions in the flat-rhombic phase lead to anomalous finite-temperature hydrogen bonding structure and dynamics.
	In bulk or conventionally hydrogen-bonded ice phases, such as the hexagonal and pentagonal monolayer phases, the most probable molecular orientations are the ones that maximize the number of hydrogen bonds~\cite{bernal1933theory}, and the protons maximize their residence time in orientations with the maximum possible hydrogen bonds.
	However, in the flat-rhombic phase, even at finite temperatures (extending past 300\,K), the most probable molecular orientations are not those that maximize hydrogen bonds but instead, those that maintain a distinct network of quasi-one-dimensional chains. \\

	To characterize the rotational motion of individual water molecules, we define $\phi$ and $\theta$ angles for molecular dipoles.
	These are depicted in Fig.~\ref{fig:angle_viz}(a) and ~\ref{fig:angle_viz}(b), respectively.
	These figures also indicate the coordinate axes used to define these angles.
	The  $\phi = \frac{\pi}{2} - \text{arccos}\left( \vb{\hat{\boldsymbol\mu}} \cdot \vb{\hat{z}} \right)$ angle captures the alignment between the molecular dipole vector $\boldsymbol\mu$ and the $z$-axis, with hats above vectors indicating normalized unit vectors.
	A water molecule with $\phi=\pm \; \pi/2$ will have its dipole vector aligned with the $\pm$ $z$-axis.
	We also wrap around the $\phi$ angles computed such that $\phi$ always lies in the range $[-\frac{\pi}{2},+\frac{\pi}{2}]$.
	The $\theta = \text{arctan}\left({\vb{c} \cdot \vb{\hat{z}}} / {\vb{c} \cdot \vb{\hat{x}}}\right)$ angle captures a molecule's rotation about its center of mass around the $y$ axis with $\vb{c} = \vb{r_{\text{OH}_\text{1}}} \times \vb{r_{\text{OH}_\text{2}}}$ and $\vb{r_{\text{OH}_\text{1}}}$ and $\vb{r_{\text{OH}_\text{2}}}$ being the OH bond vectors of an individual water molecule.
	$\theta$ measures the rotation of $\vb{c}$ around the $y$-axis.
	Since the choice of $\text{H}_1$ and $\text{H}_2$ is arbitrary, we wrap around the $\theta$ angles so that they always lie in the range $[0,\pi]$.
	\\

	To illustrate the flat-rhombic phases' apparent hesitancy to form additional hydrogen bonds, we computed the probabilities of molecular orientations defined by these $\theta$ and $\phi$ angles.
	These two-dimensional log probabilities are juxtaposed with the mean number of hydrogen bonds for each two-dimensional bin determined by the $\theta$ and $\phi$ angles, as depicted in Fig. \ref{fig:fig2}. 
	As shown in Supplementary Note V, the nanoconfined hexagonal and pentagonal phases exhibit the expected trend: molecular orientations with more hydrogen bonds should be more stable.
	In contrast, the most probable flat-rhombic molecular orientations (depicted in the left and right plots of Fig.~\ref{fig:angle_viz}(c)) correspond to only two hydrogen bonds.
	This is true even though orientations with three hydrogen bonds (like the one in the middle plot of Fig.~\ref{fig:angle_viz}(c)) are possible.
	We begin to observe these orientations with three hydrogen bonds via the entropic exploration of $\theta$ and $\phi$ angles at high temperatures and in the presence of quantum nuclear fluctuations (see Supplementary Note VI).
	However, these orientations still do not emerge as local probability maxima, suggesting they are not stable or long-lived states.
	This behavior reinforces the observation that the two-dimensional flat-rhombic ice phase prefers to remain hydrogen bonded only along one dimension, even at finite temperatures. \\

	\begin{figure}
		\includegraphics{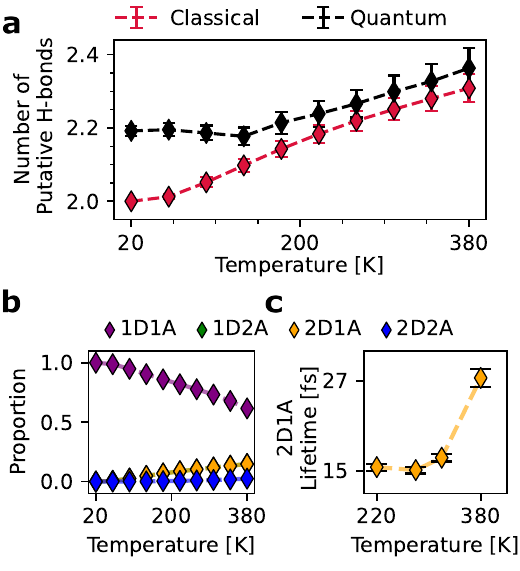}
		\caption{\textbf{$\vert$ Temperature dependence of hydrogen bonding.} a) The average number of putative/geometric hydrogen bonds across the range of temperatures in which the flat-rhombic phase is stable using classical and quantum simulations. The error bars show the standard error of the mean, as computed by block averaging over 10 blocks. b) The proportion of water molecules of each hydrogen bonding motif (defined in the text). The 1D2A proportions are perfectly aligned with the 2D1A proportions. c) The 2D1A motif's lifetime remains extremely short across the range of temperatures in which it is observed. The error bars show the standard error of the mean, computed across all instances of hydrogen bonding. Dashed lines serve as a visual guide for the eye. Source data are provided as a Source Data file.}
		\label{fig:fig3}
	\end{figure}

	The peculiar hydrogen bonding behavior of the flat-rhombic phase hence emphasizes the importance of incorporating dynamical information into the hydrogen bonding definition, as is needed in, e.g., the description of supercritical water~\cite{schienbein_supercritical_2020}.
	As shown in Fig.~\ref{fig:fig3}(a), if we employ a purely geometric definition of a hydrogen bond, counting the number of instantaneous \emph{putative} hydrogen bonds that exist on average, we observe an apparent increase in the number of hydrogen bonds with temperature.
	In the same figure, we also observe that nuclear quantum effects apparently exhibit an additional 0.2 putative hydrogen bonds, even at temperatures as low as 20 K.
	This is the result of the zero-point fluctuations lowering the energy barrier for sampling these additional hydrogen-bonded configurations, as we elaborate upon in Supplementary Note VI.
	In these configurations, hydrogen bonds form between different chains in the flat-rhombic crystal structure, as they do classically at high temperatures.
	The increased disorder due to quantum nuclear effects is not surprising, as the effect of quantum nuclear motion on properties of water has often been mapped to a temperature increase~\cite{kim_temperature-independent_2017}.
	To characterize the hydrogen bonding motifs in this structure, we introduce the notation $N$D$M$A to indicate a water molecule that donates $N$ hydrogen bonds and accepts $M$ hydrogen bonds.
	Using this notation, we see that this increase in the number of putative hydrogen bonds can be attributed to the increasing prevalence of instantaneous 2D1A motifs, i.e. instantaneous hydrogen bonds between chains. \\

	This apparent increase emerges from the shortcoming of the standard geometric definition of a hydrogen bond, which cannot distinguish fluctuating molecular orientations from long-lived orientations that correspond to true hydrogen bonds.
	As shown in Fig.~\ref{fig:fig3}(c), the lifetime of the 2D1A orientations is extremely short -- on the order of femtoseconds. 
	As suggested by Schienbein and Marx~\cite{schienbein_supercritical_2020}, these fleeting hydrogen bonds are best referred to as either putative or geometric, as they do not last long enough to be considered hydrogen bonds from a traditional perspective. 
	In Supplementary Note II, we compute the number of hydrogen bonds that survive for typical intermolecular oscillations.
	Within this definition by Schienbein and Marx~\cite{schienbein_supercritical_2020}, a hydrogen bond is not counted if its lifetime is shorter than the period of typical intermolecular oscillations.
	The resulting dynamical hydrogen bond count indeed exhibits the expected disorder-induced decrease as we raise temperature. \\

	\subsection{Dielectric nature of the flat-rhombic phase}
	
	\begin{figure}
		\centering
		\includegraphics{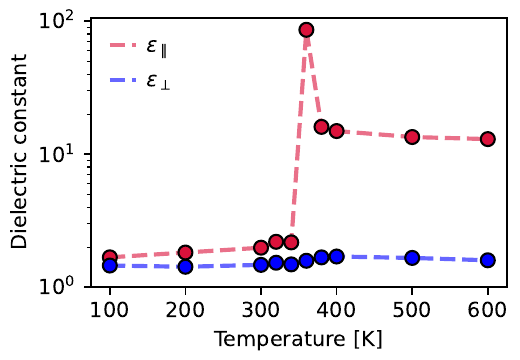}
		\caption{\textbf{$\vert$ Temperature dependence of the dielectric response.} The classical dielectric constant of the flat-rhombic phase across the range of temperatures considered in this work. The singularity at 380\,K corresponds to the transition to the hexatic phase of nanoconfined water~\cite{kapil_first-principles_2023}. Source data are provided as a Source Data file.}
		\label{fig:dielectric_response}
	\end{figure}

	A consequence of the quasi-one-dimensional hydrogen-bonded structure is that it may possess ferroelectric behavior due to its net dipole moment along the direction of the hydrogen-bonded chains. 
	Ferroelectricity in ice has been conjectured previously in force field simulations~\cite{zhao_ferroelectric_2014}, first-principles calculations of nanoconfined water~\cite{chin_ferroelectric_2021}, and experiments on confined~\cite{sugimoto_emergent_2016} and supported~\cite{su_surface-induced_1998} films. 
	To investigate the dielectric behavior of flat-rhombic ice, we model the temperature dependence of its dielectric response in Fig. \ref{fig:dielectric_response}.
	We analyzed the in-plane $\varepsilon_{\parallel}$ and out-of-plane $\varepsilon_{\perp}$ dielectric constants based on the variance of the system's polarization.
	Since our MLP only predicts the potential energy surfaces, a first principles investigation of the dielectric response would be computationally demanding due to numerous single point calculations of the electronic polarization~\cite{resta_theory_2007}.
	Therefore, for a semi-quantitative understanding, we use a simple linear polarization model based on TIP4P charges~\cite{habershon_competing_2009}. \\

	We select the TIP4P water model due to its low computational cost and a semi-quantitative description of the dielectric response of bulk~\cite{habershon_competing_2009} and confined~\cite{dufils_origin_2024} water.
	The TIP4P model underpredicts the polarization of aqueous systems as it doesn't incorporate the electronic polarization of the water molecules.
	For instance, TIP4P predicts a molecular dipole moment of 2.348 D which is 12.8\% lower compared to first principles estimates~\cite{gubskaya_total_2002}, and effectively leads to a ${\sim} 25$\% lower dielectric response.
	To incorporate the lack of the electronic polarization in the TIP4P model, we rescale the calculated dielectric response by a factor of 1.25, as has been done previously~\cite{dufils_origin_2024}.
	We report the classical dielectric response, as quantum nuclear effects are expected to only make a small quantitative difference~\cite{habershon_competing_2009}.
	\\

	We observe low and near-constant in-plane and out-of-plane dielectric constants for the flat-rhombic phase, in agreement with the experiments on nanoconfined water from Ref.~\citenum{fumagalli_anomalously_2018}.
	The low dielectric response of this phase is only an indirect outcome of vdW interactions, as they dictate the thermodynamic stability of the flat rhombic phase.
	The dielectric nature of flat-rhombic ice remains the same from 0\,K up to its phase transition into the paraelectric hexatic (disordered) water phase~\cite{kapil_first-principles_2022}.
	The temperature independence of the flat-rhombic phase's dielectric behaviors reflects the resilience of the quasi-one-dimensional structure to thermal and quantum nuclear fluctuations.
	The singularity observed in Fig. \ref{fig:dielectric_response} indicates the temperature at which the phase transition to the hexatic phase occurs.
	The hexatic phase exhibits an increased in-plane dielectric constant but a similarly small out-of-plane dielectric constant.
	\\

	\subsection{Long-ranged spatial ordering and coherent proton dynamics}
	
	\begin{figure}
		\includegraphics{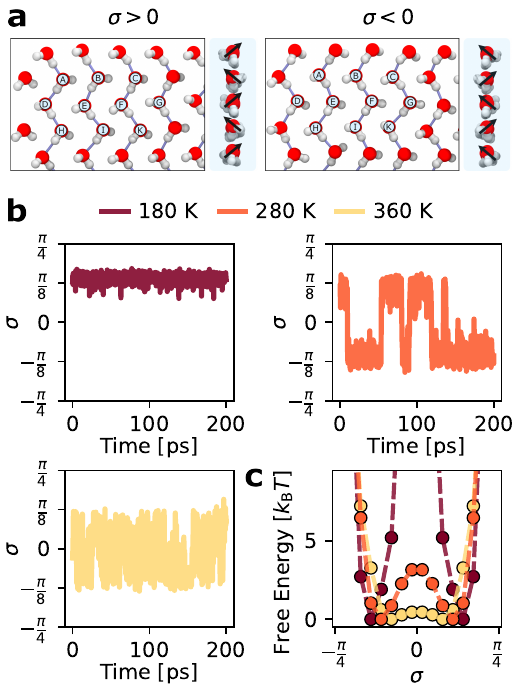}
		\caption{\textbf{$\vert$ Hydrogen bonded switching behavior.} a) The two symmetry-related structures of the flat-rhombic phase, with alternating rows of aligned dipoles,  discerned by the different signs of $\sigma$. A zero value of $\sigma$ indicates that the molecular dipoles are not aligned row-wise. The side views in each panel show the system as viewed from the right side, where the switch in the row-dependent dipole direction between the two states is clear. b) For smaller unit cells with 144 water molecules, we observe coherent switching between these two states at intermediate temperatures, as shown in the $\sigma$ trajectories for three different temperatures. c) The free energy profiles for this $\sigma$ parameter show that raising the temperature lowers the free energy barrier between these states. The smooth, low-barrier free energy profile at high temperatures indicates that the molecular dipoles lose spatial coherence. Source data are provided as a Source Data file.}
		\label{fig:fig4}
	\end{figure}

	Thus far, the results considered large simulation boxes with 576 molecules spanning tens of nanometers. 
	However, in simulation boxes containing 144 molecules in a relatively small but experimentally accessible system size on the order of nanometers~\cite{algara-siller_square_2015, zamborlini_nanobubbles_2015}, we identify unusual concerted dynamics at intermediate temperatures involving the breaking and forming of hydrogen bonds between equivalent structures. 
	In this concerted motion, we observe an exchange between the individual molecular $\phi = \pm\pi / 4$ states as a collective motion throughout the entire system, in which all the water molecules simultaneously swap the signs of their $\phi$ angles (see Supplementary Video 1). \\
	To describe this concerted motion, we define a $\sigma$ order parameter (see supporting information section for details) to distinguish the two states shown in Fig. 6(a).
	The structural change to the flat-rhombic phase between $\sigma$ values can be seen in Fig. \ref{fig:fig4}(a).
	When the sign of $\sigma$ changes, the directions of the dipoles in each row switch in a concerted fashion.
	During this motion, we also observe the complete rearrangement of the hydrogen bonding network in the lattice.
	Fig. \ref{fig:fig4}(a) shows labeled water molecules in each $\sigma$ state, illustrating that the hydrogen bonds in these two states are between different pairs of water molecules.
	At low temperatures, the system remains frozen in a single $\sigma$ value.
	The resulting $\sigma$ free energy profile contains two minima separated by a large free energy barrier, shown in Fig.~\ref{fig:fig4}(c).
	At intermediate temperatures, we begin to see exchanges between $\sigma$ signs on the order of ${\sim} 10$ ps, as shown in Fig. \ref{fig:fig4}(b).
	Compared to the low-temperature setting, the intermediate temperature free energy profile exhibits a clear finite free energy barrier between these states.
	Eventually, at high enough temperatures, the free energy profile along $\sigma$ exhibits an extremely small free energy barrier, allowing the system to explore molecular configurations freely. 
	While the highest temperatures explored here are greater than the melting temperature of ice, we believe that our trajectories correspond to a metastable state associated with the solid phase.
	Since the free energy profiles at the intermediate and high-temperature conditions are qualitatively the same, the near-free exploration of molecular configurations in the high-temperature simulations is a result of thermal activation.
	To confirm this hypothesis, we compute rotational autocorrelation functions at these different temperatures in Supplementary Note IV.

	We also note that quantum nuclear motion makes the system more disordered due to zero-point fluctuations lowering the barrier for the system's hydrogen bond network to switch between the free energy minima, as shown in Supplementary Figure 8.
	Furthermore, since nuclear quantum effects lead to increased proton disorder, the $\sigma$ parameter gets pushed closer to 0 on average.
	This causes the minima in the free energy profiles along $\sigma$ to move closer to 0 in our PIMD simulations as compared to the free energy profiles from our classical simulations.
	\\
	
	As stated earlier, this concerted motion is not observed in the larger simulation cell size that contains 576 molecules for the timescales we consider: it is only seen in simulations with 144 water molecules.
	This suggests that the free energy barrier for this concerted motion is size-extensive.
	As a result, we would expect a larger system size to either exhibit an ordered phase or a phase in which smaller domains that exhibit this behavior coexist.
	However, crystallites of nanoconfined water molecules containing around 100 water molecules are experimentally accessible, so such concerted motion could be observed in experiments as small ice crystallites forming between graphene sheets~\cite{algara-siller_square_2015} or high-pressure graphene nanobubbles formed by irradiation~\cite{zamborlini_nanobubbles_2015}. 
	Previous work has suggested that the simulation setup we employ is still reasonable for studying such encapsulated systems, even without explicit confining atoms \cite{algara-siller_square_2015,jiao_structures_2017}.
	A careful treatment of such encapsulated systems would require a detailed analysis of the water-carbon interactions at the edge of the confined water pockets.
	In this work, we ignore these edge effects and instead focus on characterizing the behavior of nanoconfined water under idealized atomistically flat nanoconfinement.
	A careful consideration of the edge effect would be an interesting and relevant topic for future work.
	\\

	\section{Discussion}

	We have analyzed the hydrogen bond structure and dynamics in the flat-rhombic phase of monolayer nanoconfined ice at first-principles-level accuracy using machine learning-driven simulations.
	Our work not only corroborates previous experimental and theoretical studies showing violations of ice rules at interfaces~\cite{jiang_first-principles_2021,lin_temperature-pressure_2023, liriano2017water, kolesnikov_quantum_2016, sugimoto_emergent_2016} but also highlights an extreme scenario where hydrogen bonding plays a limited role in stabilizing the crystalline structure of ice. 
	General ice rules must account carefully for the enthalpic penalty of breaking a hydrogen bond due to interactions with the confining walls and favourable vdW interactions. 
	Instead of maximizing the number of hydrogen bonds, flat-rhombic ice forms one-dimensional chains stabilized by vdW interactions akin to quasi-one-dimensional vdW functional materials~\cite{li2020diverse,balandin2022one}.
	The ZZ-qBI phase identified in Ref.~\citenum{lin_temperature-pressure_2023} is stable at higher lateral pressures than the flat-rhombic phase, yet it exhibits a similar behavior, indicating that such hydrogen bond topologies may occur across a broad range of nanoconfined conditions.
	This quasi-one-dimensional ice structure is robust with respect to thermal fluctuations and the rotational dynamics of water molecules, forming two or fewer hydrogen bonds up to its transition to the hexatic phase. 
	We anticipate that direct evidence of the unique structure of the flat-rhombic ice can be experimentally investigated using sum-frequency generation spectroscopy~\cite{smit_observation_2017} by disentangling the O--H vibrational stretching band~\cite{kapil_first-principles_2023} into donor-acceptor contributions. \\

	Like quasi-one-dimensional vdW functional materials that exhibit unique electronic, vibrational, and optical properties~\cite{li2020diverse, balandin2022one, zhang2019new}, the flat-rhombic phase exhibits unusual properties with technological prospects. 
	We observe long-ranged ordering of molecular dipoles along with a high spatial coherence in the hydrogen bond dynamics in the flat-rhombic ice phase, which is typically uncommon in systems containing on the order of 100s of molecules.
	This behavior is consistent with the characteristics of a two-dimensional Ising model system. 
	We anticipate that this concerted motion in flat-rhombic ice patches could be used to guide directional behaviors in nanoscale molecular devices that have already shown success in performing macroscopic-level tasks at surfaces by exploiting directed translational motion on the nanometer scale~\cite{michl2009molecular}. 
	Coherent proton dynamics observed experimentally in bulk systems, such as proton tunneling in hexagonal ice~\cite{bove_anomalous_2009} and dielectric phase transitions in molecular ferroelectrics~\cite{horiuchi_above-room-temperature_2010, horiuchi_organic_2008}, have already been computationally detected on much smaller scales~\cite{koval_ferroelectricity_2002, drechsel-grau_quantum_2014, wikfeldt_communication_2014} in flat-rhombic ice.
	This suggests that confinement effects might be crucial in phenomena like tunneling-induced or dielectric phase transitions.
	Our work opens doors to exploring new confined molecular materials with the structure-function relationships of low-dimensional functional materials that exhibit technologically relevant properties. \\
	
	\section{Methods}
	
	\subsection*{Confining Potential}
	The nanoconfined system considered in this work is a single layer of water molecules trapped between two parallel sheets -- mimicking the experimental setup in Ref.~\citenum{algara-siller_algara-siller_2015}.
	We model the interactions between the sheets and the water molecules using a simple Morse potential fit to water-carbon QMC interaction energies~\cite{chen_evidence_2016}.
	Such a potential, characterized by perfectly smooth walls, has been widely adopted in previous studies, both in force field~\cite{zhao_highly_2014, li_replica_2019, chakraborty_confined_2017} and first-principles research~\cite{corsetti_structural_2016, chen_two_2016, chen_double-layer_2017, jiang_first-principles_2021, kapil_first-principles_2022, lin_temperature-pressure_2023}.
	The uniform confinement potential has demonstrated semi-quantitative accuracy in describing the behavior of water confined within graphene-like cavities~\cite{li_replica_2019}, as corroborated by a good agreement of stable phases and melting temperatures with respect to confinement simulations that include explicit carbon atoms~\cite{jiao_structures_2017}.
	Furthermore, the uniform confinement model's atomistically flat nature allows for a clean interpretation of topological confinement effects -- a phenomenon that extends beyond the specific context of graphene-based confinement.
	
	\subsection*{Machine Learning Potential}
	For water-water interactions, we employ a newly-trained MLP~\cite{behler_constructing_2015, schran_committee_2020} at the revPBE0-D3 level~\cite{zhang_comment_1998, goerigk_thorough_2011} -- an appropriate dispersion-corrected hybrid-functional DFT level for bulk, interfacial and nanoconfined water~\cite{marsalek_quantum_2017, cheng_ab_2019, kapil_first-principles_2023, kapil_first-principles_2022}. 
	We report the training and validation protocols for the MLP in Supplementary Note IB. 
	Our simulations are run in the $NP_{xy}T$ ensemble, where $P_{xy}$ denotes the lateral pressure, or vdW pressure~\cite{algara-siller_square_2015}, that acts in the $x$ and $y$ directions.
	This lateral pressure emulates the net lateral forces on the water molecules due to the edges of the confining pocket (further details on the origins of this lateral pressure can be found in Supplementary Note IC).
	Both simulations~\cite{jiao_structures_2017} and experiments~\cite{algara-siller_square_2015} have estimated this vdW pressure to be on the gigapascal scale.
	Hence, our simulations are run at a lateral pressure of 2.0\,GPa across various temperatures within the metastability range of the flat-rhombic phase~\cite{kapil_first-principles_2022}.
	We refer the reader to Supplementary Note I for more details on the simulations \\
	
	\subsection*{First-Principles Calculations}
	To estimate the (zero temperature) static stabilization energy of a nanoconfined ice crystal or a quasi-one-dimensional chain of water molecules, we perform direct first-principles calculations using the \texttt{CP2K} code~\cite{kuhne_cp2k_2020} employing the revPBE0-D3 functional and the convergence parameters from Ref.~\citenum{kapil_first-principles_2022}. 
	We perform these calculations on geometry-optimized structures of the confined ice phases~\cite{lin_temperature-pressure_2023}.
	The geometry-optimized structures of the hexagonal, pentagonal, and flat-rhombic monolayer phases were taken from Ref.~\citenum{kapil_first-principles_2022}. 
	In contrast, the structure of the zigzag quasi bilayer ice from Ref.~\citenum{lin_temperature-pressure_2023} was optimized using the \texttt{CP2K} code~\cite{kuhne_cp2k_2020}.
	We define the stabilization energy of a crystal with respect to an isolated molecule as $E_{\textrm{lattice}} = E_{\textrm{crystal}} -  E_{\textrm{gas}}$, where $E_{\textrm{crystal}}$ is the single-point energy of the crystalline lattice per water molecule, and $E_{\textrm{gas}}$ is the single-point energy of an isolated water molecule in vacuum. 
	Similarly, we define the stabilization energy of a chain of water molecules with respect to an isolated molecule as $E_{\textrm{chain}} = E_{\textrm{chain}}^{\textrm{struct.}} -  E_{\textrm{gas}}$, where $E_{\textrm{chain}}^{\textrm{struct.}}$ is the single-point energy of the chain structure, and the stabilization energy of a crystal with respect to a chain of water molecules as $E_{\textrm{stack}} = E_{\textrm{lattice}}  - E_{\textrm{chain}}$. 
	We also estimate these quantities without the D3 dispersion correction to assess the role of vdW interactions. \\
	
	\section*{Data Availability}
	The data supporting the findings of this study are openly available on GitHub (\url{https://github.com/water-ice-group/hbond_nanoconfined_water}). This includes all simulation scripts and starting structures. Source data are provided with this paper. They are also available via this GitHub repository.
	
	\section*{Code Availability}
	All of the analysis scripts supporting the findings of this study are openly available on GitHub (\url{https://github.com/water-ice-group/hbond_nanoconfined_water}).
	
	
	\newpage

	
	\newpage
	
	\begin{acknowledgments}
		We thank Jinggang Lan, Philipp Schienbein, Michele Ceriotti, and all of A.M.'s research group members for their comments on the manuscript.
		P.R. would also like to thank David R. Reichman for his academic support during this project.
		V.K. acknowledges support from the Ernest Oppenheimer Early Career Fellowship and the Sydney Harvey Junior Research Fellowship, Churchill College, University of Cambridge.
		A.M. and X.R.A. acknowledge support from the European Union under the ``n-AQUA" European Research Council project (Grant no. 101071937).
		C.S. acknowledges partial financial support from the Deutsche Forschungsgemeinschaft (DFG, German Research Foundation) project number 500244608.
		P.R. would like to thank The Winston Churchill Foundation of the United States for their financial support.
		P.R. would also like to acknowledge that: ``This material is based upon work supported by the U.S. Department of Energy, Office of Science, Office of Advanced Scientific Computing Research, Department of
		Energy Computational Science Graduate Fellowship under Award Number DE-SC0024386. This report was prepared as an account of work sponsored by an agency of the United States Government. Neither the United States Government nor any agency thereof, nor any of their employees, makes any warranty, express or implied, or assumes any legal liability or responsibility for the accuracy, completeness, or usefulness of any information, apparatus, product, or process disclosed, or represents that its use would not infringe privately owned rights. Reference herein to any specific commercial product, process, or service by trade name, trademark, manufacturer, or otherwise does not necessarily constitute or imply its endorsement, recommendation, or favoring by the United States Government or any agency thereof. The views and opinions of authors expressed herein do not necessarily state or reflect those of the United States Government or any agency thereof.''
		We are grateful for computational support from the Swiss National Supercomputing Centre under project s1209, the UK national high-performance computing service, ARCHER2, for which access was obtained via the UKCP consortium and the EPSRC grant ref EP/P022561/1, and the Cambridge Service for Data Driven Discovery (CSD3). 
		
	\end{acknowledgments}
	
	\section*{Author Contributions}
	
	P.R., A.M., and V.K. conceived the study. P.R., V.K., and C.S. trained the machine learning potential used in this work. P.R. and X.R.A. performed molecular dynamics simulations using this MLP. All of the authors were involved in discussing results and assembling the manuscript.
	
	\section*{Competing Interests}
	
	The authors declare no competing interests.
	
	\section*{Figure Legends}
	
	\textbf{Fig. 1 $\vert$ Nanoconfined ice phases and the role of vdW stacking of hydrogen bonded chains. } a) The four nanoconfined ice phases considered in this work from Ref.~\citenum{kapil_first-principles_2022} and Ref.~\citenum{lin_temperature-pressure_2023}, along with the lateral pressures used for simulating each phase. Solid lines between water molecules indicate hydrogen bonds. We highlight chains of hydrogen-bonded water molecules in each phase. For estimating $E_{\mathrm{chain}}$, we use the unique highlighted chains in the case of the flat-rhombic and ZZ-qBI phases. For the hexagonal and pentagonal phases, we report the average value and the standard deviation as the error bar, where the standard deviation is computed across different choices of chains, as described in the text. b) For each phase, we compute the stabilization energy of a single molecule in the 0\,K crystal structure, $E_{\mathrm{lattice}}$; the (average) stabilization energy of a water molecule within the hydrogen bonded chain(s), $E_{\mathrm{chain}}$; and the stabilization energy between chains in the crystal structure, $E_{\mathrm{stack}}$. The hatched bars show the contribution of vdW interactions to the corresponding stabilization energy. Source data are provided as a Source Data file.
	
	\textbf{Fig. 2 $\vert$ Depiction of the orientations of water molecules.} a) The $\phi$ angle captures in-plane and out-of-plane fluctuations of water molecules' dipole moments. b) The $\theta$ angle captures in-plane rotations of water molecules. We refer the readers to section II.B for the mathematical definitions of the $\phi$ and $\theta$ angles. c) The left and right images depict the molecular orientations at the probability maxima in Fig.~\ref{fig:fig2}(a), while the middle image depicts the molecular orientation at the maximum for the number of putative hydrogen bonds in Fig.~\ref{fig:fig2}(b).
	
	\textbf{Fig. 3 $\vert$ Molecular orientation and the number of putative hydrogen bonds.} a) The two-dimensional log probabilities along the $\phi$ and $\theta$ parameters for the flat-rhombic phase at 2 GPa and three different temperatures. b) The average number of putative hydrogen bonds associated with each $(\theta,\phi)$ molecular orientation for the flat-rhombic phase under the same conditions. Relevant molecular orientations are depicted in Fig.~\ref{fig:angle_viz}(c). We employ the geometric definition of the hydrogen bond from Ref.~\citenum{luzar_hydrogen-bond_1996}. Source data are provided as a Source Data file.
	
	\textbf{Fig. 4 $\vert$ Temperature dependence of hydrogen bonding.} a) The average number of putative/geometric hydrogen bonds across the range of temperatures in which the flat-rhombic phase is stable using classical and quantum simulations. The error bars show the standard error of the mean, as computed by block averaging over 10 blocks. b) The proportion of water molecules of each hydrogen bonding motif (defined in the text). The 1D2A proportions are perfectly aligned with the 2D1A proportions. c) The 2D1A motif's lifetime remains extremely short across the range of temperatures in which it is observed. The error bars show the standard error of the mean, computed across all instances of hydrogen bonding. Dashed lines serve as a visual guide for the eye. Source data are provided as a Source Data file.
	
	\textbf{Fig. 5 $\vert$ Temperature dependence of the dielectric response.} The classical dielectric constant of the flat-rhombic phase across the range of temperatures considered in this work. The singularity at 380\,K corresponds to the transition to the hexatic phase of nanoconfined water~\cite{kapil_first-principles_2023}. Source data are provided as a Source Data file.
	
	\textbf{Fig. 6 $\vert$ Hydrogen bonded switching behavior.} a) The two symmetry-related structures of the flat-rhombic phase, with alternating rows of aligned dipoles,  discerned by the different signs of $\sigma$. A zero value of $\sigma$ indicates that the molecular dipoles are not aligned row-wise. The side views in each panel show the system as viewed from the right side, where the switch in the row-dependent dipole direction between the two states is clear. b) For smaller unit cells with 144 water molecules, we observe coherent switching between these two states at intermediate temperatures, as shown in the $\sigma$ trajectories for three different temperatures. c) The free energy profiles for this $\sigma$ parameter show that raising the temperature lowers the free energy barrier between these states. The smooth, low-barrier free energy profile at high temperatures indicates that the molecular dipoles lose spatial coherence. Source data are provided as a Source Data file.
	
	

	\clearpage
	\newpage
	\setcounter{section}{0}
	\twocolumngrid
\onecolumngrid
\begin{center}
    \Large \bfseries Supplementary Information:\\
    Quasi-one-dimensional hydrogen bonding in nanoconfined ice
\end{center}
\vspace{1em} 
\twocolumngrid
	
	\section{C\lowercase{omputational} D\lowercase{etails} \label{appendix:comp_details}}
	
	\subsection{Simulation model of nanoconfined water}
	
	The system considered in this work is a set of water molecules trapped between two parallel confining sheets, as shown in Supplementary Figure~\ref{fig:simulation_setup}. 
	The two parallel confining sheets are kept 5.0\,\AA{} apart, measured as the distance between the centers of the atomic nuclei in each confining sheet. 
	This results in the water molecules forming a single monolayer parallel to each confining sheet.
	
	\begin{figure}[h]
		\includegraphics{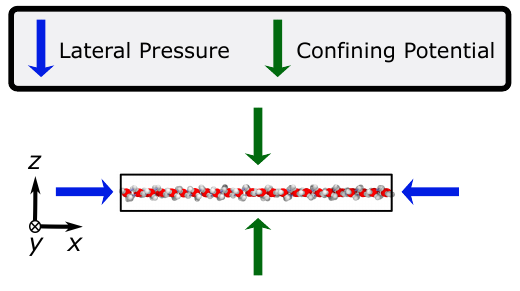}
		\caption {\textbf{$\vert$ Simulation setup overview.} Schematic overview of the confined system and the directions of the lateral pressure and confining potential.}
		\label{fig:simulation_setup}
	\end{figure}

	The total potential energy of this system is written as a sum of the potential energy of the water molecules and a confining potential~\cite{truskett_thermodynamic_2001,zhao_ferroelectric_2014,  corsetti_structural_2016, chen_two_2016, chakraborty_confined_2017, li_replica_2019, jiang_first-principles_2021, kapil_first-principles_2022}. 
	The confining potential is a Morse potential fitted to Quantum Monte Carlo water-carbon total energies used in Refs.~\citenum{chen_two_2016, chen_double-layer_2017, kapil_first-principles_2022}. 
	The confining potential is uniform in the plane of confinement and depends explicitly on the perpendicular distance of a water molecule from the confining sheets. 
	The total potential energy of the water molecules is represented using a high-dimensional Behler-Parrinello neural network potential~\cite{behler_constructing_2015} fitted to total energies and forces calculated at revPBE0-D3 density functional theory level~\cite{zhang_comment_1998, goerigk_thorough_2011}. \\
	
	\subsection{Machine Learning Potential}
	\label{appendix:MLP}

	To train the machine learning potential, we employ an active learning protocol called query by committee (QbC) implemented by Schran et al.~\cite{schran_committee_2020}. 
	In the QbC approach, the MLP committee is initially trained on a small set of structures. 
	Subsequent structures are added to the training set from a larger dictionary only if they have large committee disagreements. 
	By only adding structures with large committee disagreements to our training set, we can iteratively and systematically build up a compact training set of informative structures. 
	Previous work used the bulk and interfacial water potential from Ref.~\citenum{schran_committee_2020} as a starting point and supplemented additional monolayer and bulk structures at a variety of temperatures and lateral pressures~\cite{kapil_first-principles_2022}. 
	In this work, we further added monolayer, bilayer, trilayer, and tetralayer structures from simulations of confining widths of 6.5, 9.0, 9.5, and 11.5\,\AA{}.
	We also included structures from PIMD simulation trajectories so that the training dataset includes the new configurations that are accessible with the inclusion of nuclear quantum effects.
	The average training error of the resulting MLP used in this work is 83.82 meV/\AA{} for forces and 3.38 meV/water for energies. \\
	
	To assess the neural network's ability to generalize to unseen configurations, we compared the energies and forces produced by the MLP to those produced by the reference revPBE0-D3 functional on new structures that were not seen during training time.
	These structures were generated by running 2 nanosecond $NP_{xy}T$ simulations at 2.0 GPa starting from each of the solid phases identified in Ref.~\citenum{kapil_first-principles_2022}.
	Each solid phase was simulated at 100 K, 300 K, and 600 K.
	For some of these simulations, the solid phase was metastable, and so the trajectory remained in the solid phase.
	For other simulations, the solid phase was unstable, and the water molecules in the simulation entered a disordered phase or glassy phase (depending on the conditions).
	Hence, these trajectories contained a diverse set of ordered and disordered configurations that span a broad range of local molecular configurations.
	We computed the errors between the MLP's predictions and the reference revPBE0-D3 calculations over all of the configurations generated in each of these trajectories.
	The resulting mean absolute errors (MAE) for the energies and root-mean-squared errors (RMSE) for average atomic forces are shown in Supplementary Table~\ref{tab:mlp_errors}.
	These errors are similar in magnitude to the errors of MLPs employed in other studies~\cite{kapil_first-principles_2022,lin_temperature-pressure_2023}, which have been validated against reference DFT simulations.
	We also note that this model~\cite{kapil_first-principles_2023} (and its predecessors based on the same machine learning architecture~\cite{kapil_inexpensive_2020, shepherd_efficient_2021}) have been validated against experimental IR, Raman and sum frequency generation spectra, demonstrating an excellent description of real-time dynamics of aqueous systems.
	\\
	
	\begin{center}
		\begin{table}[t!]
			\caption{\textbf{$\vert$ Force field validation.} Average MLP errors on configurations that were not seen during training time. These configurations comes from $NP_{xy}T$ simulations at 2.0 GPa covering a wide range of newly-generated ordered and disordered structures.}
			\begin{ruledtabular}
				\begin{tabular}{ || c || c | c ||}
					& Energy MAE & Avg Atomic Force \\
					Temperature & [meV/water] & RMSE [meV/\AA{}] \\
					\hline
					100 K & 5.8 & 67.6 \\
					\hline
					300 K & 4.8 & 81.8 \\
					\hline
					600 K & 5.6 & 103.4 \\
				\end{tabular}
			\end{ruledtabular}
			\label{tab:mlp_errors}
		\end{table}
	\end{center}
	
	\subsection{vdW pressure
		\label{appendix:vdW}}
	
	An important component of this system is the vdW pressure (or lateral pressure) experienced between the sheets of the confining material. 
	For instance, if the confining material is graphene, the attractive vdW forces between the two graphene sheets will pull them closer together. 
	If a pocket of water is encapsulated between these two sheets, the graphene sheets will begin to close inwards, pushing the water pocket into an increasingly small region of space. 
	This will continue until the internal pressure of the water balances this attractive vdW force. 
	This results in an effective lateral pressure from the termination of the pocket that acts in the directions perpendicular to the confining potential~\cite{algara-siller_square_2015}. 
	We model this lateral pressure by running simulations in the $NP_{xy}T$ ensemble.
	
	\subsection{Molecular dynamics simulations}

	The molecular dynamics simulations were performed using \texttt{i-PI}~\cite{kapil_i-pi_2019} with the \texttt{n2p2}~\cite{singraber_library-based_2019} code interfaced with \texttt{LAMMPS}~\cite{thompson_lammps_2022} to perform MLP energy and gradient calculations. 
	Molecular dynamics simulations used cells containing $144$ and $576$ molecules. 
	To sample the $NP_{xy}T$ ensemble, we employed a flexible barostat~\cite{martyna_molecular_1999} (constrained to only allow lateral cell fluctuations) with a time constant of 1000\,fs and optimally-damped generalized Langevin equation thermostats~\cite{ceriotti_langevin_2009} to control the temperature of the physical and the barostat degrees of freedom. 
	To calculate dynamical properties, we performed simulations in the $NVT$ ensemble using a stochastic velocity rescaling thermostat~\cite{bussi_canonical_2007} with a time constant of 100\,fs. 
	Using a BAOAB splitting of the isothermal isobaric Liouville operator~\cite{leimkuhler_robust_2013, kapil_modeling_2019}, we used a timestep of 1.0\,fs to run simulations for 1\,ns.
	
	\subsection{Confining Potential}
	
	As discussed earlier, we used an implicit confining potential to emulate the interactions between the water molecules and the confining material.
	The two parallel walls are kept at a fixed distance of 5\,\AA{} away from each other, and the nanoconfined water molecules all lie between these two implicit walls.
	If a water molecule is a distance $z$ away from one of the walls, the potential that it experiences is a Morse potential of the form:
	\begin{eqnarray}
		V_{morse} = 
		D_0 \qty[\qty(1 - e^{-a\qty(z-z_0)})^2 - 1]
	\end{eqnarray}
	with the parameter values $D_0 = 5.78 \cross 10^{-2}$ eV, $z_0 = 3.85$ \AA{}, and $a = 0.92\text{ \AA{}}^{-1}$.
	These values were obtained from a fit to water-carbon Quantum Monte Carlo (QMC) interaction energies~\cite{chen_evidence_2016}.

	\subsection{Path integral molecular dynamics simulations}

	The path integral molecular dynamics (PIMD) simulations used the same software and parameters as the molecular dynamics simulations. 
	We employed simulation cells containing $144$ molecules with 32 imaginary time slices.  
	We used an optimally-damped generalized Langevin equation thermostat to control the barostat temperature and a local path-integral Langevin equation thermostat~\cite{ceriotti_efficient_2012} to thermalize the physical degrees of freedom. 
	Using a BAOAB splitting of the path-integral isothermal isobaric Liouville operator, we used a timestep of 0.5\,fs to run simulations for 1\,ns.
	For structural and thermodynamic quantities (e.g., free energy profiles), we average results over all the imaginary time slices from the PIMD simulations.
	For dynamical quantities, we use the centroids of the imaginary time slices.
	
	\subsection{Computing free energy profiles}
	
	We used a simple binning procedure to compute free energy profiles along a particular variable, $ x$.
	Given a molecular dynamics trajectory, we assemble a list of all values that the variable $x$ takes on in each frame of the trajectory.
	We then compute the probability histogram $P(x)$ with an appropriately chosen number of bins.
	The free energy profile in units of $k_B T$ is then reported as $F(x) = - \log(P(x))$, where $k_B$ is Boltzmann's constant.
	To account for quantum nuclear effects, we perform the binning procedure on the trajectory of the individual imaginary time slices of our PIMD simulations and average the histograms.
	
	\section{H\lowercase{ydrogen} B\lowercase{ond} D\lowercase{efinition} \label{appendix:orient}}
	
	We employ the geometric hydrogen bond definition of Luzar and Chandler~\cite{luzar_hydrogen-bond_1996}. 
	Two water molecules are considered to be (putatively) hydrogen-bonded if they satisfy the following two conditions: (1) The distance between the two oxygen atoms is less than 3.5\,\AA{}, and (2) the angle formed by the O-O vector and one of the O-H bond vectors is less than $30^{\circ}$. 
	The sensitivity of the hydrogen bonding trends for the flat-rhombic phase with respect to the parameters of this geometric definition are shown in Supplementary Figure~\ref{fig:geometric_sensitivity}. \\
	
	In Fig.~\ref{fig:fig3}(a) of the main text, we showed that employing such a geometric hydrogen bond definition resulted in the number of hydrogen bonds seeming to increase as temperature is raised.
	This contradicts the expected behavior from a standard energy-entropy trade-off perspective.
	We suggested that this was because this geometric definition still included fleeting hydrogen bonds that only lasted for times shorter than intermolecular timescales.
	Employing the definition by Schienbein and Marx~\cite{schienbein_supercritical_2020}, we also computed the number of dynamical hydrogen bonds as a function of temperature.
	Here, a dynamical hydrogen bond is one that satisfies the above geometric criteria for at least 0.2 ps, the typical timescale for intermolecular oscillations for monolayer ice as well as supercritical and room temperature water~\cite{schienbein_supercritical_2020}.
	The resulting dynamical hydrogen bond count exhibits the expected decreasing trend, as shown in Supplementary Figure~\ref{fig:dynamical_hbond_count}.
	
	\begin{figure}
		\centering
		\includegraphics{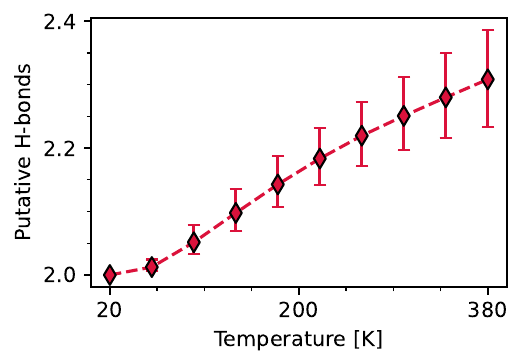}
		\caption{\textbf{$\vert$ Geometric hydrogen bond definition.} The number of (putative) hydrogen bonds in our classical simulations of the flat-rhombic phase as a function of temperature. The main curve is the same as in the classical (red) curve of Fig. \ref{fig:fig3}(a) in the main text. The upper/lower error bounds are obtained by increasing/decreasing the two hydrogen bonding criteria in the text by 5\% each.}
		\label{fig:geometric_sensitivity}
	\end{figure}

	\begin{figure}
	\centering
	\includegraphics{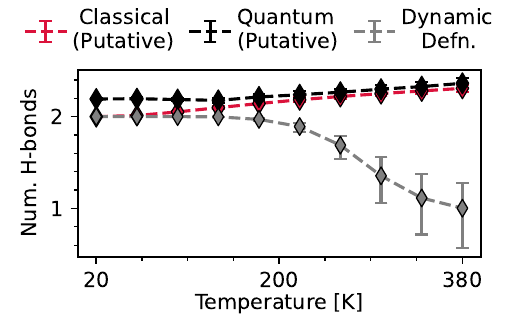}
		\caption{\textbf{$\vert$ Dynamical hydrogen bond count.} The grey data points show the number of hydrogen bonds that survive at least 0.2 picoseconds, with the lower and upper error bounds given by a 0.3 and 0.1 picosecond cutoff, respectively. The red and black lines are the same as in Fig. \ref{fig:fig3}(a) of the main text, showing the number of geometric/putative hydrogen bonds as a function of temperature. The error bars for the putative lines show the standard error of the mean, as computed by block averaging over 10 blocks.}
	\label{fig:dynamical_hbond_count}
	\end{figure}
	
	\section{$\sigma$ O\lowercase{rder} P\lowercase{arameter} D\lowercase{efinition}}
	
	To describe the concerted motion shown in Fig.~\ref{fig:fig4}(a) of the main text, we define a $\sigma$ order parameter to distinguish the two states.
	The side views of Fig.~\ref{fig:fig4}(a) show that the individual dipole directions of each row of water molecules change when the flat-rhombic phase exchanges between these states.
	We use $\phi_{ij}$ to denote the $\phi$ angle of the water molecule in row $i$ and column $j$ of the flat-rhombic lattice.
	We define the switching parameter $b_i$ that alternates signs between rows and the $\sigma$ order parameter as
	\begin{eqnarray}
		b_i =
		\begin{cases}
			+1 &\text{odd $i$} \\
			-1 &\text{even $i$}
		\end{cases}
		\\
		\sigma = \frac{1}{n} \sum_{i,j} b_i \phi_{ij},
		\label{eq:PR_OP}
	\end{eqnarray}
	where $n$ is the number of water molecules in our system.
	The above equation sums over the product of the row-specific $b_i$ parameter and the molecule-specific $\phi_{ij}$ angle between the geometric dipole of the water molecule and the confinement plane.
	With these definitions, the $b_i \phi_{ij}$ terms in the summation will have the same sign if the rows of water molecules have alternating dipole directions.
	Therefore, the absolute value of this order parameter $|\sigma|$ should take on large values for the crystal structure of the flat-rhombic phase, and the motion shown in Fig.~\ref{fig:fig4}(a) should correspond to a sign change in this $\sigma$ order parameter.
	Given the arbitrary choice of positive values for odd $i$ and negative values for even $i$ in the definition of $b_i$, only relative changes in the sign of $\sigma$ should be considered, as the absolute sign of $\sigma$ has no meaning.
	Furthermore, note that during the concerted motion described above, the value of the $\sigma$ order parameter will switch its sign very quickly.
	If the motion of the protons is not correlated, then the values of the $\phi_{ij}$ values will be uncorrelated with the row index $i$.
	This will cause the $\sigma$ value to take on intermediate values near 0, which is indicative of a loss of proton order.
	\\
	
	\section{R\lowercase{otational} A\lowercase{utocorrelation} F\lowercase{unctions}}
	\label{appendix:rotational_legendre}
	
	\begin{figure}
		\centering
		\includegraphics{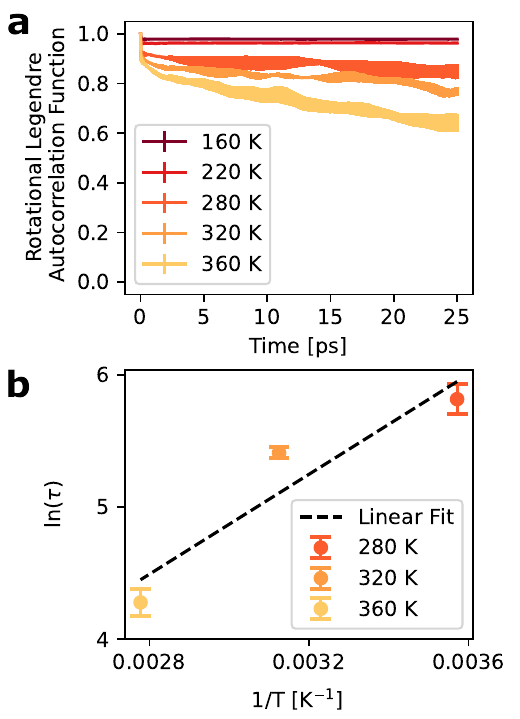}
		\caption{\textbf{$\vert$ Rotational Legendre Autocorrelation Functions.} a) The rotational Legendre autocorrelation functions of water molecules in the flat-rhombic phase at various temperatures. These are computed as described in Ref.~\cite{wilkins2017nuclear} with $n=1$. These are averaged over 5 different trajectories. The width of each line is the error, as computed by the standard error of the mean across the trajectories. b) The temperature dependence of the rotational relaxation time $\tau$, fit to an Arrhenius form (dashed black line). We estimate $\tau$ as the integral of the above autocorrelation functions, and we only consider temperatures where the concerted motion is observed. The error bars for $\ln(\tau)$ are computed by propagating the errors from the top panel.}
		\label{fig:legendre}
	\end{figure}
	
	For dynamical insights into the rotational motion of water molecules, we estimate the rotational Legendre autocorrelation functions of the O--H bond with $n=1$, following the procedure described in Ref.~\cite{wilkins2017nuclear}.
	This autocorrelation function will decay to 0 as the initial rotational configuration of the water molecule relaxes to a random orientation. 
	For each displayed temperature in Supplementary Figure~\ref{fig:legendre}(a), we estimate the autocorrelation function over a 25 picosecond window, averaged over 5 trajectories.
	The definite integral of this function, from $0$ to $\infty$, gives the rotational relaxation time. \\
	Supplementary Figure~\ref{fig:legendre}(a) clearly shows two different types of rotational motion in the flat-rhombic phase. 
	As we mentioned in the discussion surrounding Fig.~\ref{fig:fig2} of the main text, water molecules in low-temperature conditions remain localized in their molecular orientations up to the simulation time considered in this work. 
	At low temperatures such as 160 and 220\,K, this behaviour leads to a sharp initial decay in the rotational Legendre autocorrelation plots in Supplementary Figure~\ref{fig:legendre}(a).
	This sharp decay is due to thermal fluctuations within the $(\theta, \phi)$ free energy minimum.
	As a result, these autocorrelation functions exhibit a nearly infinite rotational relaxation time, since the water molecules are stuck in the free energy minimum that they begin in.
	The diverging relaxation time is an artefact of the absence of molecular rotation sampling within the timescales of our simulations.
	At intermediate temperatures such as 280\,K and 360\,K, we observe finite rotational relaxation times, consistent with the $(\theta, \phi)$ exploration of water molecules in Fig.~\ref{fig:fig2} of the main text.
	We also note a good fit of the rotational relaxation times between 280\,K and 360\,K to an Arrhenius equation, suggesting a common activation barrier dictating the rotational motion. 
	We do not include the temperatures 160 and 220\,K in the fit, as at these temperatures we do not sample the relevant relaxation event, implying an $\infty$ relaxation time.
	We do not account for quantum nuclear motion as they are expected to only make a quantitative difference to this trend~\cite{wilkins_accurate_2019, habershon_competing_2009}. 
	Specifically in the context of the flat-rhombic phase, the correspondence between Fig.~\ref{fig:fig2} of the main text and Supplementary Figure~\ref{fig:legendre}, and the quantitative increase in the exploration of $(\theta, \phi)$ orientations due to quantum nuclear effects (see Supplementary Figure~\ref{fig:nqe_2D}) implies a systematic increase the relaxation time.
	
	\section{H\lowercase{ydrogen} B\lowercase{onding} C\lowercase{haracterization of the} H\lowercase{exagonal and} P\lowercase{entagonal} P\lowercase{hases} \label{appendix:HB_hex_pent}}
	
	\begin{figure}
		\centering
		\includegraphics{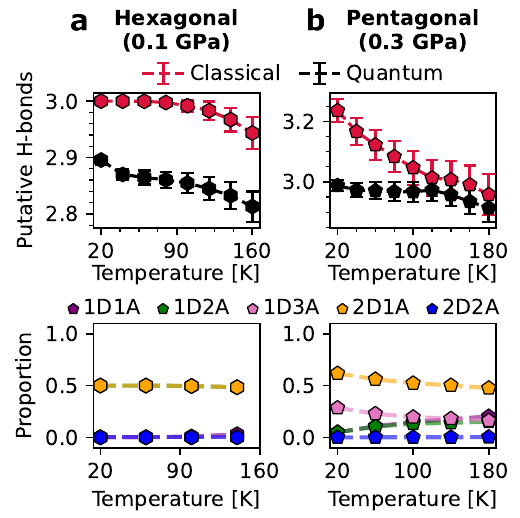}
		\caption{\textbf{$\vert$ Hydrogen bonding in the hexagonal and pentagonal phase.} a) Temperature dependence of the number of geometric/putative hydrogen bonds in the hexagonal and pentagonal nanoconfined ice phases. b) The proportion of water molecules occupying each $N$D$M$A state for the hexagonal and pentagonal ice phases. The error bars for the top panels show the standard error of the mean, as computed by block averaging over 10 blocks.}
		\label{fig:nqe_nH}
	\end{figure}
	
	As shown in Supplementary Figure.~\ref{fig:nqe_nH}, the hexagonal and pentagonal phases exhibit decreasing geometric hydrogen bond counts with increasing temperature. This trend is expected based on bulk ice's conventional hydrogen bonding behavior. As seen from the $N$D$M$A  analysis, in the hexagonal phase, the decline in the number of hydrogen bonds with temperature is due to a slight reduction in the population of the 1D2A / 2D1A states forming 3 hydrogen bonds per molecule and a simultaneous increase in 1D1A states forming 2 hydrogen bonds per molecule. In the pentagonal phase, the decline in the number of hydrogen bonds is due to a reduction in the population of the 2D1A and 1D3A motifs, forming 3 and 4 hydrogen bonds per molecule, respectively, and a simultaneous increase in the population of 1D1A and 1D2A motifs, forming 2 and 3 hydrogen bonds per molecule. Thus, thermal fluctuations disrupt the $N$D$M$A  states that form the (locally) highest possible number of hydrogen bonds. \\
	
	\begin{figure}
		\centering
		\includegraphics{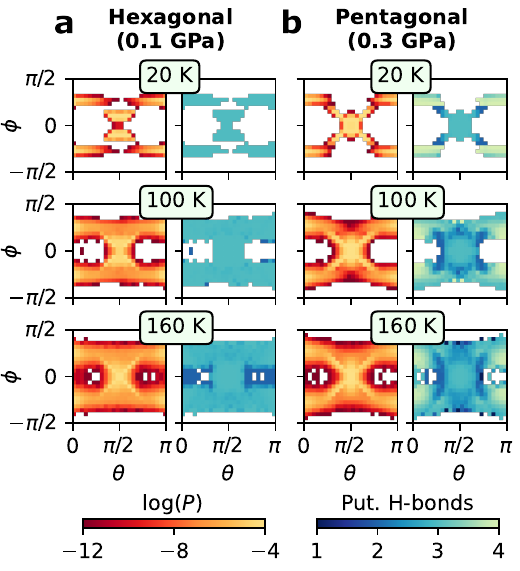}
		\caption{\textbf{$\vert$ Molecular orientations in the hexagonal and pentagonal phases.} a) The log-probability and number of hydrogen bonds in $(\theta,\phi)$-space for the hexagonal phase. b) The same plots for the pentagonal phase.}
		\label{fig:hp_2Ds}
	\end{figure}
	
	In Fig. \ref{fig:fig2} of the main text, we provide 2D log-probability profiles in $\phi$ and $\theta$ space for molecules in the flat-rhombic phase. We also provide the average number of hydrogen bonds observed in the flat-rhombic phase in $\phi$ and $\theta$ space. Surprisingly, the maxima in the log-probability profiles do not align with the maximum in the number of hydrogen bonds plot. This suggested that the number of hydrogen bonds alone does not predict the stability of a nanoconfined ice configuration. Supplementary Figure~\ref{fig:hp_2Ds} shows the same types of plots for the hexagonal and pentagonal phases. These plots serve as a baseline for our expectations from bulk ice, where the Bernal-Fowler ice rules would dictate that bulk ice configurations that satisfy the maximum hydrogen bonding coordination of 4 would be the most stable ordered configurations. The plots in Supplementary Figure \ref{fig:hp_2Ds} for the hexagonal and pentagonal phases exhibit exact alignment between the log-probability maxima and the number of hydrogen bond maxima, which suggests that this hydrogen bonding-oriented picture holds for these phases. \\
	
	\section{I\lowercase{mpact of} Q\lowercase{uantum} N\lowercase{uclear} E\lowercase{ffects on the} F\lowercase{lat-}R\lowercase{hombic} P\lowercase{hase} \label{appendix:NQE_flat_rhombic}}
	
\begin{figure}
		\centering
		\includegraphics{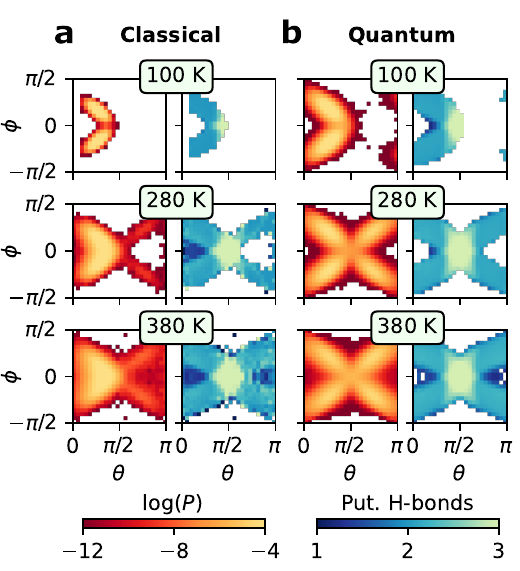}
		\caption{\textbf{$\vert$ Nuclear quantum effects and molecular orientations.} a) The classical log-probability profiles and number of hydrogen bond plots in $(\theta,\phi)$-space. b) The same plots but computed for PIMD simulation trajectories that account for quantum nuclear motion.}
		\label{fig:nqe_2D}
	\end{figure}

	Supplementary Figure~\ref{fig:nqe_2D}(b) shows the log-probability and average number of hydrogen bonds in $(\theta,\phi)$-space for the PIMD simulations. Supplementary Figure~\ref{fig:nqe_2D}(a) just serves as a convenient replication of the results from Fig.~\ref{fig:fig2}(b) of the main text for comparison. The major result, i.e., the fact that the log-probability maxima do not align with the maximum in the number of hydrogen bond plot, holds in the presence of quantum nuclear motion, suggesting that hydrogen bonds are not the primary stabilization interaction in the flat-rhombic phase. However, the PIMD simulations exhibit a broader exploration of $(\theta,\phi)$-space at all temperatures. In particular, they enhance the in-plane rotational motion of the molecules about its dipole vector. \\
	
	We also compute the free energy profile along $\sigma$ in these PIMD simulations in Supplementary Figure \ref{fig:nqe_sigma_fep}. Comparing Supplementary Figure~\ref{fig:nqe_sigma_fep} to Fig.~\ref{fig:fig4}(c) of the main text shows that the classical and PIMD simulations exhibit the same qualitative behavior. However, the free energy barriers for the $\sigma$-switching motion are lower due to zero-point fluctuations, and the minima in the free energy profiles move closer to $\sigma=0$ due to the increased proton disorder, as discussed in the main text.
	
	\begin{figure}
		\centering
		\includegraphics{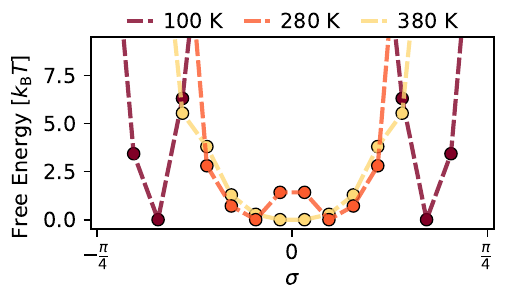}
		\caption{\textbf{$\vert$ Nuclear quantum effects and the $\sigma$ order parameter.} Free energy profiles for the $\sigma$ order parameter computed from PIMD simulations at various temperatures. This can be directly compared to Fig. \ref{fig:fig4}(c) in the main text.}
		\label{fig:nqe_sigma_fep}
	\end{figure}

	
	%

	\newpage

\end{document}